\documentstyle[aps,multicol,epsfig]{revtex}
\draft
\def\vec#1{{\rm\bf #1}}
\begin{document}

\title{Exponential Decay of Correlations in a Model for Strongly
Disordered 2D Nematic Elastomers}

\author{Y.-K. Yu\cite{byline} and P.L. Taylor}
\address{Department of Physics, Case Western Reserve University,
Cleveland, Ohio 44106-7079, USA}
\author{E.M. Terentjev}
\address{
Cavendish Laboratory,
University of Cambridge, Madingley Road, Cambridge,
CB3 0HE, U.K.}
\maketitle


\begin{abstract}
Lattice Monte-Carlo simulations were performed to study the equilibrium
ordering in a two-dimensional nematic system with quenched
random disorder. When the disordering field, which competes against the
aligning
effect of the Frank elasticity, is sufficiently strong, the
long-range correlation of the director orientation is found to decay as a
simple
exponential,
$e^{-r/\xi}$.  The correlation length $\xi$ itself also decays exponentially
with increasing strength of the disordering field.   This result represents
a new type
of behavior, distinct from the Gaussian and power-law decays predicted by
some theories.
\end{abstract}


\begin{multicols}{2}

The effects of quenched random disorder are crucial for physical
systems as diverse as random ferromagnets, vortex lattices in type-II
superconductors, structural glasses, foams and liquid crystals.  It is
therefore of
great importance to attain an understanding of the way in which the range
of the
ordering of such systems varies as a function of the relative strengths of the
ordering and disordering fields.  In this Letter we present results from a
study of
one simple model system.  The behavior we observe is qualitatively
different from
any that has previously been discussed.

Strong disorder, manifesting itself in stochastic
values of certain microscopic parameters, does not permit the system to
achieve its
ordered ground state at low temperatures. This results in a variety of
glass-like
structures with only short-range order, but with correlations distinctly
different
from those in a high-temperature liquid state. An example is the
Sherrington-Kirkpatrick spin glass
\cite{she,doz}, which is represented theoretically by an Ising model with
random exchange  constants,
the variance of whose fluctuations acts as a measure of the degree of
disorder.
When the characteristic energy of the disordering field
is small,
or the number density of discrete random impurities is low, the system is
capable
of establishing ideal ordering on a small but macroscopic length scale.
On larger scales the cooperative effects of the random field may prevail over
the  demand for uniform thermodynamic order, and the long-range correlations
decrease with distance in systems with sufficiently low space
dimensionality. The
possibility then exists of separating the effects of elasticity, which on a
continuum
level penalizes the  deviations from the ground state order, from those of the
continuum, coarse-grained,  weak random field.

The concept of a well-established local order that is distorted
on length scales larger than a characteristic coherence length $\xi_d$ to form
textures with  zero global
average requires a multi-component order parameter. The simplest example
is the vector order parameter of ferromagnets, in which a dilute random
array of impurities gives rise to  correlated spin-glass textures
\cite{ima,chu}.   A system that is in many ways analogous to the
ferromagnet
is a  nematic liquid crystal having a uniaxial tensor order
parameter, where the principal axis (the
director {\bf n}) can  form birefringent  textures. Nematic elastomers  are
very close in their behavior to the correlated spin glasses, with network
crosslinks acting as local sources of
weak quenched disorder \cite{fri}. A recent analysis based on the
concept of hierarchical replica symmetry breaking \cite{gia} has sugested that
at long distances a quasi long-range order in vortex arrays
pertains, {\it i.e.} the fluctuations diverge logarithmically and
a simple power-law decay of long range correlations is predicted.  This is in
contrast to the general arguments  of Imry and Ma \cite{ima}, who predict
that the
long-range correlation of a continuous order parameter should decay as $\exp
\left[  -(r/\xi_d)^{4-d}\right] $ with $d$ the dimensionality of space.

All the existing continuum models are based on the notion of
weak disorder, which is difficult to control parametrically. The
region of applicability of such a concept could be tested experimentally,
if a series of similar systems with systematically increasing random
field were available. In reality, however,  random magnets exist
only with a limited range of possible impurities \cite{magexp}, and it is in
general
difficult to monitor the spatial variation of local magnetization in
textures. Nematic elastomers offer more promise for experimental
studies because the birefringent textures are readily accessible to
several standard optical techniques \cite{stu}. This gives us the motivation to
study such systems in depth.

In this Letter we use the direct lattice Monte-Carlo simulation of equilibrium
textures in a model corresponding to  nematic
elastomers, which thus shows strong similarity to spin glasses. We look
for
the change in long-range correlations that occurs in  response to a
variation of
random field strength. Because we need to
analyze large systems, for computational efficiency we consider only the case
of $d=2$. The nematic
director field {\bf n}({\bf r}) is assumed to be of constant  (unit) amplitude
and its direction to vary in the plane so that a single angle $\theta(x,y)$
is the only coordinate to enter the Hamiltonian. This corresponds to a
locally
well-established orientational order forming a large-scale  texture, and
is closely related to a 2D XY model ($d=2,n=2$), the difference lying in the
invariance of the nematic elastomer to inversion of the director
 ${\bf n}$ to $-{\bf n}$.

  We consider  a nematic with misoriented rod-like crosslinks,
which locally impose a preferred orientation direction on the director field.
This is analogous to  a continuum ferromagnet with randomly
distributed impurities.  The Hamiltonian of such a system can be written as:
\begin{eqnarray}
{\cal H} = \int
 \frac{1}{2}K
\left(\nabla  \vec{n}\right)^2{\bf dr}
- \frac{1}{2} \sum_{\bf \ell} \gamma
\left(\vec{k}_{\bf \ell} \cdot \vec{n} \right)^2.
 \label{Hram}
\end{eqnarray}
Here $K$ is the Frank elastic constant (in the single-constant approximation)
and $\gamma$ the strength of
the randomly oriented field, whose local direction is defined by the unit
vector $\vec{k}_{\bf \ell}$ at the crosslink location ${\bf \ell}$.
In a
typical elastomer the crosslinks are separated by distances much smaller
than the
expected correlation length $\xi_d$, and typically of order 1--5 nm.

To perform the Monte Carlo study we first make the assumption that the
essential element of the disorder lies in the orientational randomness of the
crosslinks rather than their spatial distribution.  This allows us to use a
model in
which the sites $\bf{\ell}$ form a regular lattice, each site of which
contains a
random field
$\bf k(\ell)$, with no
correlations between neighbors. The  density of crosslinks, $\rho_0$,
is thus assumed  uniform.

The
minimization of the free energy (\ref{Hram})  determines in each cell the
local
director orientation  $\bf n(r_\ell)$,
which at zero temperature will be a function only of the dimensionless
parameter
$D^* \equiv 4\gamma /K$. The resulting discrete dimensionless
Hamiltonian, scaled by the Frank constant $K$, takes the form
\begin{eqnarray}
{\cal H} = \sum_{\bf \ell} \Bigl[ -\sum_{\bf \delta}
\cos^2 \left(\theta_{\bf \ell}-\theta_{\bf \ell + \delta}\right) -
\frac{1}{2} D^*
\left(\vec{k}_{\bf \ell} \cdot \vec{n}_{\bf \ell} \right)^2 \Bigr] ,
 \label{Hdis}
\end{eqnarray}
where  ${\bf \delta}$ are the
nearest-neighbor lattice vectors, and the lattice spacing, $\rho_0^{-1/2}$,
becomes
the unit of length.

The key information we seek is the form of the correlation function
$C({\bf r})$ that describes the distance over which the director
remains largely unchanged in orientation.
In the systems we shall study, in which the director is confined to a
2-dimensional  manifold, this requires computation of the average over
positions
$\bf r_0$ of $C({\bf r})
=2\langle \cos^2[ \theta ({\bf r}+{\bf r_0}) -\theta({\bf r_0})] \rangle -1$.
For a macroscopically isotropic (polydomain) system we
expect this quantity to be a function only of $|\bf{ r}|$, and so we
have chosen to evaluate it as the average over all directions of $\bf{r}$.

 Rather than perform the
simulations on a very large system we studied lattices of modest size
(typically $60 \times 60$) and performed many repeated evaluations of
$C(\vec{r})$ for different realizations of the random array of crosslink
orientations. In this way it was possible
to estimate the reliability of the results from the width of the
spread in $C(\vec{r})$ produced.  The effect of the lattice size was found
by comparing
the results for lattices of $60 \times 60$ and $40 \times 40$ sites.  The
values of
$C(\vec{r})$ were indistinguishable for small $|\vec{r}|$, and differed by
only a few per
cent even when $|\vec{r}|$ was half the size of the smaller lattice.

The procedure adopted to find the minimum energy configuration was to
run the Metropolis
algorithm for the Monte Carlo method at a finite temperature and then
to reduce the temperature to zero for the remainder of the run. This
process of annealing and quenching avoided having the system be stranded
in some of the smaller local energy minima. Some further tests of the
robustness of the results were made by
varying the initial conditions. A starting point of complete alignment,
in which the directors at each lattice site were initially mutually
parallel, produced results for
$C(\vec{r})$ that tended to be larger than when the starting point was a
completely random set of orientations. Each series of runs was then
continued until the results from the
two opposing starting points converged onto each other.

A more physically significant choice to be made concerned the
boundary conditions used.
The evaluation of a reliable form for $C(\vec{r})$ is greatly facilitated
by using periodic boundary conditions, since then all lattice points are
equivalent. On the other hand, the
absence of any physical boundary to the system has the consequence
that topological defects cannot evaporate at a surface, and can only be
eliminated by a process of mutual
annihilation. The weakness of the effective attraction that occurs
between certain types
of defects hinders this process, and makes some defects very long
lived. The presence of a free boundary thus greatly speeds the process
of equilibration, as defects are able
simply to wander off the edge of the system. The disadvantage that
comes with the use of free boundaries, however, is the fact that not
all lattice sites are
equivalent, and the calculation of correlation functions must be
limited to pairs of sites that are in the interior of the system.
In the simulations to be described below, it was
indeed found that the use of free boundary conditions led to a
greatly accelerated approach to equilibrium, and so this scheme was
generally employed.

The decision to adopt free boundary conditions was reinforced by an
examination of the consequences of applying a boundary condition
intended to represent the effects of
surface anchoring. In this modification, a uniform field was applied
only to the lattice sites lying at the boundaries, and the simulation
was run with the two opposing
initial configurations, the totally random and the completely aligned.
It was then found that the two sets of
correlation functions never approached each other, no matter how long
the run was continued, the initially
aligned state giving a larger value for the correlation function at
all separations {\bf r} than
the one starting from totally random director field. To avoid
this effect we have preferred the free boundary conditions.

The correlation function $C(r)$ was calculated as the average of
 $2({\bf n}_{\ell_0}\cdot {\bf n}_{\ell_0+ \ell})^2-1$ over all
sites ${\bf \ell}$
contained within a circular shell of radius $r$,
centered on site ${\bf \ell}_0$ and of thickness one lattice spacing.
The
calculations thus describe the effects of the
competition between the random field, which tends to destroy all
correlations, and the
nematic interaction, which tends to introduce long-range correlations
in the director
orientation.

\begin{figure}[h]
\centerline{ \epsfxsize=7cm \epsfbox{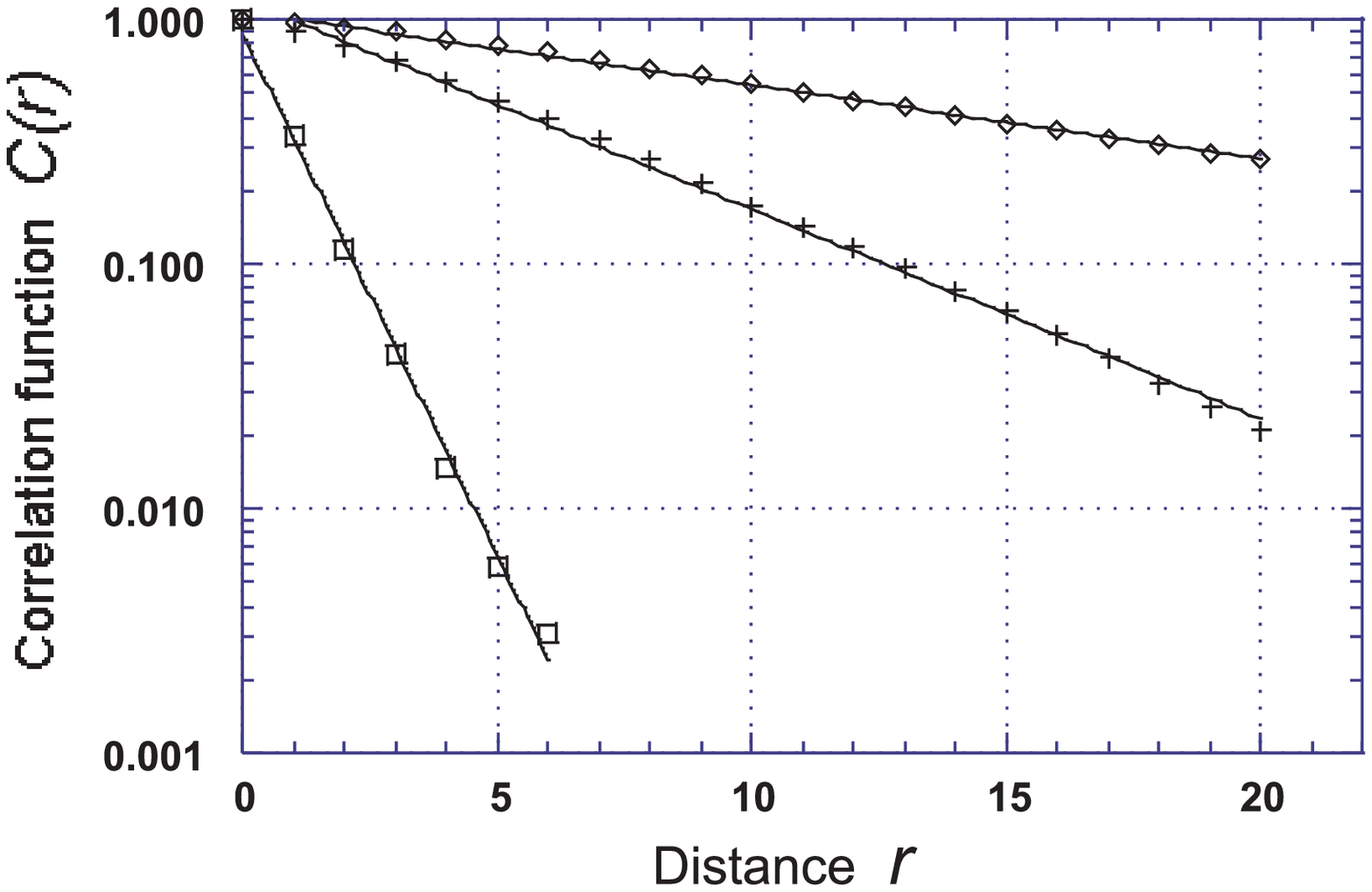}}   
\label{Yu1}     
\end{figure} 
\vspace{-0.4cm}
\parbox[t]{8cm}{\scriptsize FIG.1 \ \ The
orientation correlation  function $C(r)$ is shown as a function of distance,
measured in lattice spacings, for values of $D^*$, the relative
strength of the random field due to crosslinks, equal to 0.5 (diamonds),
1.0 (crosses), and 3.0 (squares).}   
\vspace{0.3cm}

When $D^*$ is large, the effect of the random fields of the
crosslinks dominates, and
the decay of $C(r)$ is extremely rapid. As $D^*$ is reduced, the
range of the correlation function gradually increases. In the range
studied, for which $0.5 < D^* <3.5$, the correlation function is found
to approximate fairly closely a simple exponential.
This is illustrated in Fig.~\ref{Yu1}, which shows on a semilogarithmic
plot the form of $C(r)$ for $D^* = 0.5$, 1.0, and 3.0 with lines representing
the exponential fit.

The necessity of checking the robustness of the calculation by
choosing two opposite
starting configurations was confirmed by performing an
average of 50 runs of 75 000 time steps at each of the 3600 sites,
starting from both
ordered and random initial states. The two results were in
 disagreement by about 30\% at large $r$.  Only after the runs had been
continued to 150 000 time
steps did the two curves approach each other sufficiently to be
indistinguishable.

\begin{figure}[h]
\centerline{ \epsfxsize=7cm \epsfbox{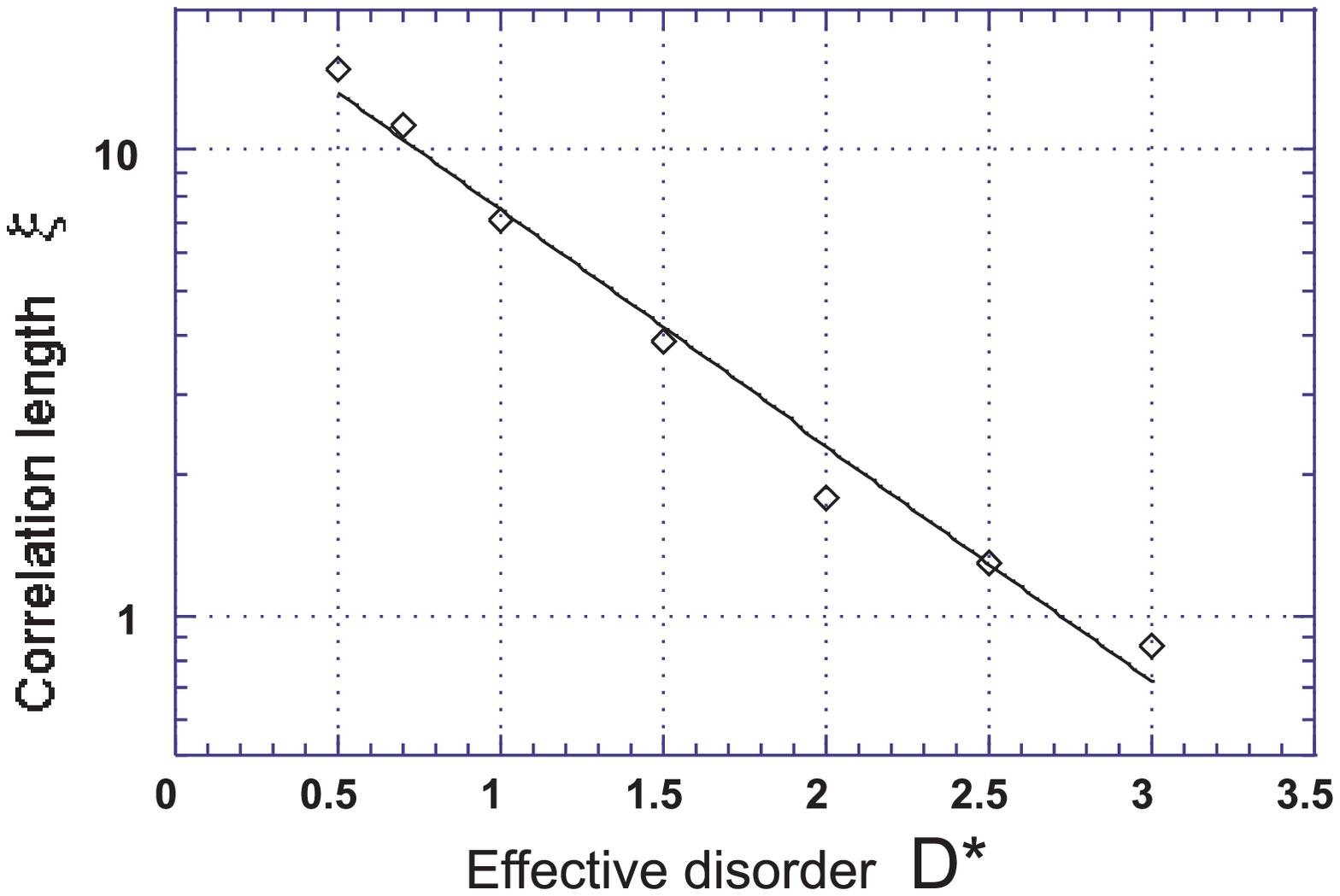}}   
\label{Yu3}     
\end{figure}

The linearity of the plots in Fig.~\ref{Yu1} tells us that
the  correlation function can be written in the form $C(r) = e^{-r/\xi}$
over the range of $D^*$ studied.  This form for the correlation function
permits us to identify $\xi(D^*)$ as the correlation length for this
system, measured in units of the lattice spacing, $\rho_0^{-1/2}$.  We
expect it to be related to the ``domain size'' \cite{ima}.  When we examine
the variation of $\xi(D^*)$ with $D^*$ we find that this in turn is fitted
by an
exponential to within the  precision of our computations, as is shown in
Fig.~\ref{Yu3}, and  thus is of the form
$\xi(D^*) \propto e^{-D^*/D_0}$. This result is somewhat surprising in
view of the fact that $\xi$ must tend to infinity as
$D^*$ vanishes, and suggests that there are two distinct regimes for
the correlation behavior in this randomly disordered nematic system.
It is unfortunate that the amount of computation needed to
produce reliable results increases very steeply as $D^*$ is reduced
below 0.5, and that
it was consequently not feasible to perform simulations to follow the
form of $\xi(D^*)$ down into this small-$D^*$ crossover region.  At the
other extreme,
where $D^*$ becomes larger than 3, the correlation length diminishes to a
fraction of a
lattice spacing.  (This statement is not as devoid of meaning as it
appears, since it is
merely a way of saying that the correlation in orientation between nearest
neighbors has
fallen to a number less than $e^{-1}$.)

\vspace{-0.4cm}
\parbox[t]{8cm}{\scriptsize FIG.2 \ \ The
effective correlation length $\xi$ is found to vary exponentially with
the strength of disorder $D^*$.}   
\vspace{0.3cm}

Further calculations were made for a non-zero applied external field,
which enters the  Hamiltonian as
$-({\bf h}\cdot {\bf n_\ell})^2$ and introduces an additional dimensionless
parameter
$(h^2/K)$ in Eq.(\ref{Hdis}). The results here were less unexpected, in
that the
correlation function $C(r)$, after an initial rapid decay, approached
asymptotically
a constant value that reflects the susceptibility of the system and was,
accordingly, proportional to $(h^2/K)^2$ for small $h$.

To summarize, the results of our simulation describe the equilibrium
ordering of nematic elastomers,
whose large-scale alignment is disturbed by the quenched misorienting
effect of
network crosslinks. The continuum coarse-grained model used to describe this
system is closely similar to that for a spin glass. We considered the
case of a
two-dimensional XY model, which is a good description of a thin flat sample of
nematic elastomer
with the director confined to this plane by planar boundary conditions.
This is the typical experimental arrangement
described earlier. Accordingly, the model falls into the class of
$(d=2,n=2)$ continuum Hamiltonians with the quenched random field weak enough
to allow formulation in terms of a continuum displacement field
$\theta(x,y)$.

The classical arguments of Imry and Ma \cite{ima} predict correlations to
decay as
$C(r) \sim \exp [-(r/\xi_2)^2]$, while a treatment accounting for
replica symmetry breaking \cite{gia}
 predicts $C(r) \sim (r/\xi)^{-x}$, with $x$ a positive fractional
exponent.  Our
results appear to be quite unambiguous and to provide a totally different
correlation law:
$C(r) \approx \exp ( -r/\xi)$ with the characteristic decay length $\xi \simeq
23\exp (-1.16 D^*)$ whenever $D^* > 0.5$.  This behavior cannot persist down to
vanishingly weak disorder, as it fails to predict an infinite value for
$\xi$ when $D^*
\rightarrow 0$.  This suggests  that there is a crossover to another
regime at smaller values of $D^*=\gamma /K$
than considered here, and which might be within
the range of validity of the analytical predictions \cite{ima,gia}. The
fact that an
extrapolation of our data to $D^* = 0$ gives a decay length roughly equal
in size to the
linear dimensions of the lattice used in the calculation is almost
certainly only a
coincidence; the large-$D^*$ behavior from which the extrapolation is made
should be independent of the lattice size.

We are left with a number of unanswered questions.  What is the physical
explanation of
the exponential form of the decay length?  Over what range of disorder
strength does it
persist?  Is this result unique to two dimensions or does it have a more
general
applicability?  The resolution of these problems will require some deeper
insight and a
significant dedication of computational resources.

This work was made possible by NATO Collaboration Grant CRG941041, and was
supported by the NSF ALCOM Science and Technology Center
 under Grant DMR89-20147 (PLT and YKY) and by EPSRC UK (EMT).

\end{multicols}
\end{document}